\documentclass{elsart}
\usepackage{graphicx}
\begin{document}

\begin{frontmatter}

\title{Isospin dependence of isobaric ratio Y($^{3}$H)/Y($^{3}$He) 
and its relation to temperature}

\author{M. Veselsky\thanksref{Contact}\thanksref{SASc}},
\author{R. W. Ibbotson\thanksref{BNL}},
\author{R. Laforest\thanksref{MIR}},
\author{E. Ramakrishnan\thanksref{MuCal}},
\author{D. J. Rowland\thanksref{MIR}},
\author{A. Ruangma},
\author{E. M. Winchester},
\author{E. Martin} and
\author{S. J. Yennello}

\address{Cyclotron Institute, Texas A\&M University, 
College Station, TX 77843-3366, USA} 

\thanks[Contact]
{Phone: (979)-845-1411, fax: (979)-845-1899, e-mail: veselsky@comp.tamu.edu}
\thanks[SASc]
{On leave of absence from Institute of Physics of SASc, Bratislava, Slovakia}
\thanks[BNL]
{Present address: Brookhaven National Laboratory, Brookhaven, New York, USA}
\thanks[MIR]
{Present address: Mallinckrodt Institute of Radiology, St. Louis, Missouri, USA}
\thanks[MuCal]{Present address: Microcal Software Inc, One Roundhouse Plaza, 
Northampton, MA 01060, USA}

%\maketitle

\begin{abstract}
{
A dependence of the isobaric ratio Y($^{3}$H)/Y($^{3}$He) on the N/Z ratio 
of the reconstructed quasiprojectile for the reaction of $^{28}$Si 
beam with $^{112,124}$Sn targets at two different projectile energies of 30
and 50 MeV/nucleon is presented. We demonstrate a linear dependence 
of the observable ln(Y($^{3}$H)/Y($^{3}$He)) 
on the N/Z ratio of the quasiprojectile and show the dependence 
of the slope on the reconstructed excitation
energy of the quasiprojectile. We relate this slope 
dependence at a given excitation energy to the temperature of the fragmenting
system. Using the model assumptions 
of the statistical multifragmentation model, a method of temperature 
determination is proposed. A caloric curve is constructed 
and compared to the result of the double isotope ratio method for the 
same set of the data and to the results of other studies. 
}
\end{abstract}


%\pacs{25.70.Mn, 25.70.Pq}

\end{frontmatter}

%\maketitle

%\section{Introduction}

Multifragmentation of highly excited nuclei has been subject 
of significant interest for many years. Since the nuclei are 
two-component systems consisting of protons and neutrons, the influence 
of isospin on the disassembly of hot nuclei is a question of general 
interest. Experimental studies have demonstrated the influence of the isospin 
of projectile and target on the isotopic yields of fragments 
\cite{wada87,yenn94,john96,john97,ramak}. 
Theoretical framework for fragmentation of asymmetric systems 
have been developed \cite{lamb,MuSe,chomaz,bali,baran,barz88} 
and a possible phase transition into an isospin-symmetric liquid and 
an isospin-asymmetric gas phase have been suggested. 
In our recent work \cite{veselsky} we presented the characteristics of 
the quasiprojectiles with $Z=12-15$, reconstructed from the fully isotopically 
resolved fragments with $ Z_{f}\leq 5 $, detected in the forward angles using 
the FAUST multidetector array \cite{faust} at Cyclotron Institute of Texas 
A\&M University. The distributions of the quasiprojectile velocity, mass, 
charge and excitation energy were obtained in the reaction of a $^{28}$Si beam 
with $^{112,124}$Sn targets at projectile energies 30 and 50 
MeV/nucleon. We demonstrated that these experimental observables 
of the quasiprojectile can be reproduced satisfactorily by hybrid 
simulations, using a code implementing the model of deep inelastic transfer 
\cite{tassan-got} for a calculation of the properties of 
excited quasiprojectiles and the SMM code \cite{bondorf-smm} 
for a description of multifragmentation of excited quasiprojectiles. Further 
details can be found in \cite{veselsky}. The level of an agreement obtained 
implies that the influence of the non-statistical effects caused by violent 
collisions of the projectile and the target on the properties of detected 
quasiprojectiles can be practically excluded for the selected events. 
The obtained set of data allows 
further study of thermal quasiprojectile multifragmentation, especially 
the study of the influence of the quasiprojectile isospin on the properties 
of fragmenting system. 
This data is of specific interest because the isospin of the system 
which actually undergoes multifragmentation is known with good precision.  
The first part of the study was presented in our recent work \cite{isodist} 
where multiplicities and N/Z ratios of light charged particles \hbox{( LCPs )} 
and intermediate mass fragments ( IMFs ) were studied. Different trends of 
multiplicities and an inhomogeneous distribution of isospin 
between light charged particles and intermediate mass fragments were observed. 
The multiplicity of LCPs increases rapidly with proton excess, while 
the multiplicity of IMFs increases slowly with increasing neutron number. 
For quasiprojectiles with large proton excess an inhomogeneous 
distribution of isospin into proton-rich LCPs and more symmetric IMFs occurs. 
Furthermore, dependences of the isobaric ratio Y($^{3}$H)/Y($^{3}$He) 
on the N/Z ratio ( as a variable closely connected to isospin ) and 
on the excitation energy of the reconstructed quasiprojectiles were 
presented in \cite{isodist}. In this work we will present a detailed study of 
the trends of the isobaric ratio Y($^{3}$H)/Y($^{3}$He). The relationship 
of Y($^{3}$H)/Y($^{3}$He) to thermodynamical observables will be discussed. 

%\section{Isospin dependence of isobaric ratio Y($^{3}$H)/Y($^{3}$He)}

\begin{figure}[!tbp]
\centering
\vspace{10mm}
\includegraphics[height=13cm]{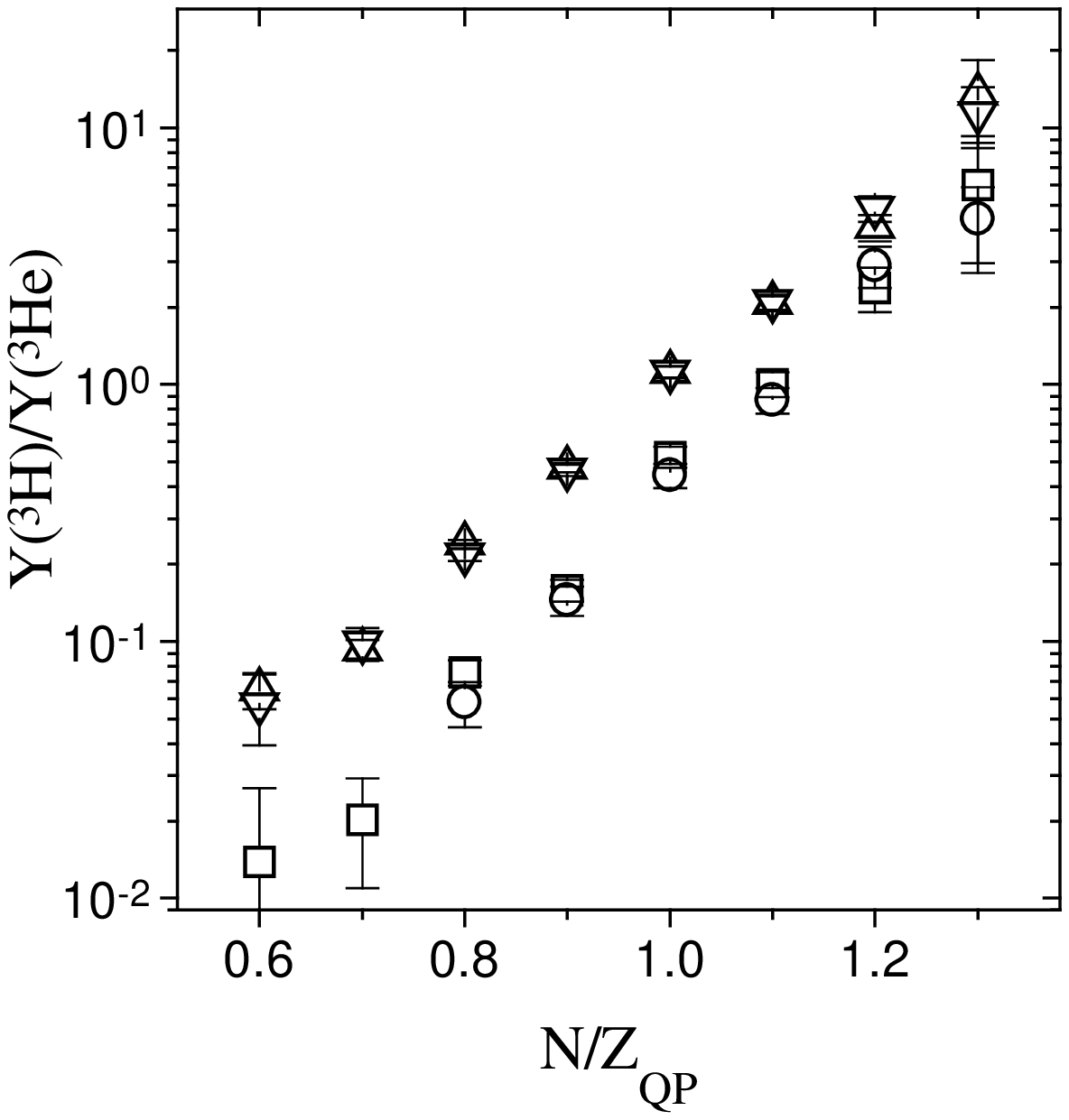}
\vspace{10mm}
\caption{ 
Dependence of the isobaric ratio 
Y($^{3}$H)/Y($^{3}$He) on the N/Z ratio of the fully 
isotopically resolved quasiprojectiles with $Z_{tot}=12-15$.
Squares - $^{28}$Si(30MeV/nucleon)+$^{112}$Sn, 
circles - $^{28}$Si(30MeV/nucleon)+$^{124}$Sn, 
up triangles - $^{28}$Si(50MeV/nucleon)+$^{112}$Sn, 
down triangles - $^{28}$Si(50MeV/nucleon)+$^{124}$Sn.
}
\label{fig1}
\end{figure}

The dependences of the isobaric ratio 
Y($^{3}$H)/Y($^{3}$He) on the N/Z ratio of the quasiprojectile 
\hbox{( $N/Z_{QP}$ )} are shown in Fig. \ref{fig1} 
for the reaction of a $^{28}$Si beam with $^{112,124}$Sn targets 
at projectile energies of 30 and 50 MeV/nucleon. 
The quasiprojectiles with $Z$ = 12 - 15, reconstructed from the fully 
isotopically resolved charged fragments with $ Z_{f}\leq 5 $, 
are selected ( for details of the experiment see \cite{veselsky,laforest} ). 
Various symbols represent the experimental data for different targets 
at each beam energy. 
The data within each set form nearly straight lines thus implying an 
exponential dependence of the isobaric ratio Y($^{3}$H)/Y($^{3}$He) 
on $N/Z_{QP}$. 
For a given projectile energy, the Y($^{3}$H)/Y($^{3}$He) 
dependence on $N/Z_{QP}$ is practically identical for the two different 
targets $^{112,124}$Sn. This shows that the isotopic trends 
are determined by the isospin of the excited system which fragments and 
the trends obtained by comparison of the data for different projectile-target 
systems describe only the overall trends. 
A detailed discussion of this issue is in our previous work \cite{isodist}. 
For the two different projectile energies 
30 and 50 MeV/nucleon the Y($^{3}$H)/Y($^{3}$He) dependence 
exhibits a different slope. In the above 
mentioned work \cite{veselsky}, the distributions of the apparent excitation 
energy ( $E^{*}_{app}$ ) are practically identical 
for the different targets at a given projectile energy and shift 
to the increasing values of $E^{*}_{app}$ with an increasing beam energy. 
The mean values of $E^{*}_{app}$ are 101.2 and 102.3 MeV for $^{112}$Sn 
and $^{124}$Sn targets at the projectile energy 30 MeV/nucleon and 142.8 MeV 
for both targets at the projectile energy 50 MeV/nucleon. Such a 
similarity may imply that the slope of the dependence of isobaric 
ratio Y($^{3}$H)/Y($^{3}$He) on $N/Z_{QP}$ correlates 
with the mean excitation energy of the quasiprojectile. 
Preliminary indications of this dependence have been reported 
in ref. \cite{isodist}. This assumption may 
be examined by extracting the isobaric ratio Y($^{3}$H)/Y($^{3}$He) 
for different bins of the apparent excitation energy 
of the quasiprojectile. 

\begin{figure}[!tbp]
\centering
\vspace{10mm}
\includegraphics[height=13cm]{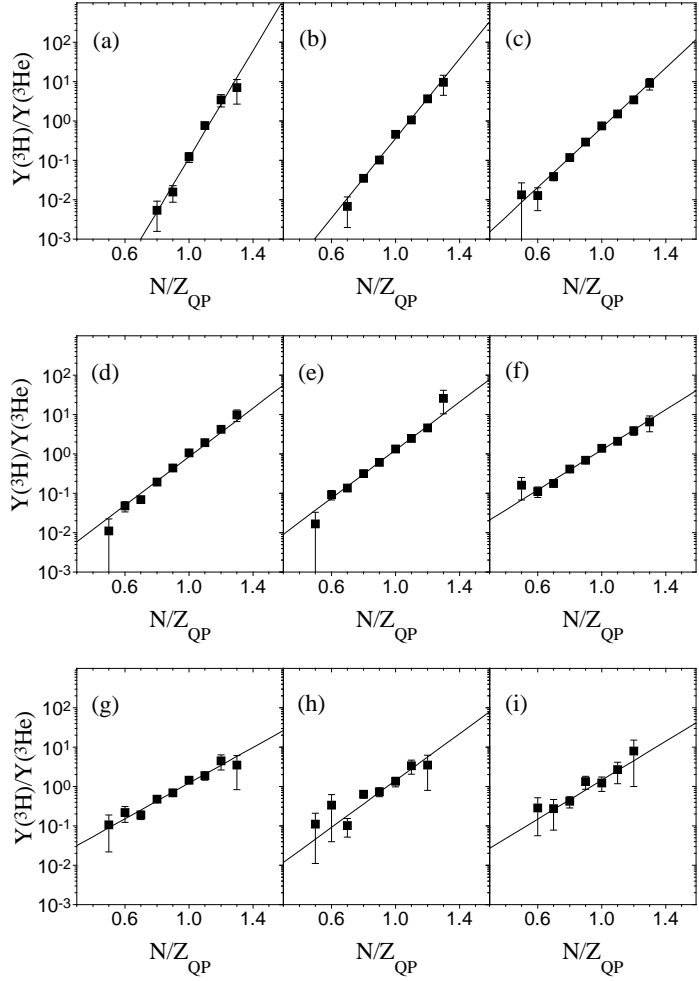}
\vspace{10mm}
\caption{ 
Dependence of the isobaric ratio 
Y($^{3}$H)/Y($^{3}$He) on the N/Z ratio of the fully 
isotopically resolved quasiprojectiles with $Z_{tot}=12-15$ for 
the different bins of $\epsilon_{app}$. 
The panels (a) - (i) correspond to the excitation energy 
bins given in Table \ref{table1}. 
The full set of the data from reactions  
of $^{28}$Si beam with $^{112,124}$Sn targets at 
the projectile energies 30 and 50 MeV/nucleon was used. 
}
\label{fig2}
\end{figure}

Fig. \ref{fig2} shows the dependences of the isobaric 
ratio Y($^{3}$H)/Y($^{3}$He) on $N/Z_{QP}$ for the nine bins of the apparent 
excitation energy per mass unit of the quasiprojectile ( $\epsilon_{app}$ ). 
The data for both targets and projectile energies were combined 
to increase the statistics. 
This is possible because for a given excitation energy bin, the 
Y($^{3}$H)/Y($^{3}$He) dependences agree within 
the statistical deviations. The squares represent the experimental data and 
the lines are the fits obtained using the MINUIT minimization 
package \cite{minuit}. 
As one can see on Fig. \ref{fig2}, linearity is the overall feature 
of the logarithmic plots of the Y($^{3}$H)/Y($^{3}$He) ratio in all 
excitation energy bins and is especially significant in the excitation 
energy bins with a high statistics. The slopes are steepest at low 
excitation energies and become flatter with increasing excitation energy. 

Since the experimental dependences of the isobaric ratio Y($^{3}$H)/Y($^{3}$He)
on $N/Z_{QP}$ correlate with $\epsilon_{app}$ 
and are linear on a logarithmic scale ( and thus depend exponentially 
on $N/Z_{QP}$ ), it would be of interest to relate the slope 
to the thermodynamical observables of the quasiprojectile multifragmentation. 
As a starting point we chose the statistical multifragmentation model 
\hbox{( SMM )} \cite{bondorf-smm} which was employed in the simulation 
successfully describing our data \cite{veselsky,isodist}. 
In the macrocanonical approach, the SMM gives the mean number 
of the emitted fragments of a given $ N_{k} $ and $ Z_{k} $ as 

\begin{equation}
 <M_{N_{k}Z_{k}}>=g_{N_{k}Z_{k}}\frac{V_{f}}{\lambda _{T}^{3}}A^{3/2}
 \exp (\frac{F_{N_{k}Z_{k}}(T,V)-\mu _{n}N_{k}-\mu _{p}Z_{k}}{T}) 
\label{multnz}
\end{equation}

\noindent 
where $T$ is the freeze-out temperature, $V$ is the freeze-out volume, 
$ g_{N_{k}Z_{k}} $ is the degeneracy of the fragment ground state, $ V_{f} $ 
is the free volume available for translational motion of the fragment, 
$ \lambda _{T} $ is the thermal wavelength of the nucleon, 
$ F_{N_{k}Z_{k}}(T,V) $ is the internal free energy of the fragment and 
$ \mu _{n} $ and $ \mu _{p} $ are the chemical potentials for neutrons 
and protons. The assumption of chemical equilibrium is not incorporated into 
SMM \cite{bondorf-smm} and the chemical potentials $ \mu _{n}$ and $\mu _{p} $ 
are used as a Lagrange multipliers ensuring the conservation of the 
mean neutron and proton number. When considering 
the isospin dependence of the experimental Y($^{3}$H)/Y($^{3}$He) 
ratio for a given excitation energy bin, the exponential dependence suggests 
a possibility to determine a single temperature for each bin as 
a simplest possible interpretation. The isobaric ratio Y($^{3}$H)/Y($^{3}$He) 
then can be determined as 

\begin{equation}
 \frac{Y(^{3}H)}{Y(^{3}He)}=\frac{<M_{21}>}{<M_{12}>}
 =\exp{(-\frac{F_{21}(T,V)-F_{12}(T,V)-\mu _{n}+\mu _{p}}{T})} 
\label{r3}
\end{equation}

\noindent 
where we used the fact that $ V_{f} $ and $ \lambda _{T} $ are the same for
the whole partition and that the ground state spins of both fragments are equal.

We approximate the $ \mu _{n}-\mu _{p} $ by the difference in the 
neutron ( $ S_{n} $ ) and proton ( $ S_{p} $ ) ground state separation energy. 
According to the model of a nuclear Fermi gas at zero temperature, $ S_{n} $ 
and $ S_{p} $ can be identified with the chemical potentials for protons 
and neutrons in the ground state. When introducing finite temperature, the 
Fermi gas model predicts only a moderate change of the chemical potential 
for temperatures well below the Fermi energy. Thus the approximation 
of chemical potentials by separation energies can be still 
considered valid for highly excited thermally equilibrated nuclei 
and determines the correct chemical potentials prior to multifragmentation. 
During multifragmentation the system expands which creates two different 
trends. When considering the system as Fermi gas expanding 
to the freeze-out density ( typically 0.4$\rho_0$ for the present case 
according to the formula used in the SMM ), 
the values of the Fermi energy decrease by more than 40 \%. 
The same behavior can be expected also 
for $ \mu _{n}-\mu _{p} $. On the other hand, chemical potentials of 
nucleons are influenced by the process of multifragmentation. 
As reported in our previous work \cite{isodist}, the IMFs 
are more isospin-symmetric than light charged particles. Within 
statistical multifragmentation model, the isobaric ratio Y($^{3}$H)/Y($^{3}$He) 
can be identified with the ratio of mean densities of free neutrons and 
protons for a given fragment partition ( as one can see 
after dividing equation \ref{multnz} by freeze-out volume ). As follows from 
Figs. \ref{fig1},\ref{fig2}, such ratios differ dramatically from the starting 
values and one can expect also the values of $ \mu _{n}-\mu _{p} $ 
significantly higher than those of an expanded Fermi gas. 
When assuming that both effects have similar magnitude, the approximation used 
can be considered as a reasonable estimate of the actual values. 
For the full set of data used in Fig.\ref{fig2}, $ \mu _{n}-\mu _{p} $ 
can be approximated as 

\begin{equation}
\mu _{n}-\mu _{p} \approx S_{p}-S_{n}=(-43.44\pm 0.97)+
(38.57\pm0.94)N/Z_{QP} \hbox{ [MeV]. }
\label{bnbp}
\end{equation}

\noindent 
Experimental mass excesses \cite{wapstra} have been used for 
the evaluation. The dependence of $ S_{n}-S_{p} $ on the N/Z ratio of 
the quasiprojectile can be considered linear with good precision. 
The difference in the internal free energy of the fragments $^{3}$H 
and $^{3}$He, according to the evaluation formulas given in \cite{bondorf-smm}, 
is of the order of a few hundred keV on a broad range of $N/Z_{QP}$ and 
changes much more slowly than $ \mu _{n}-\mu _{p} $.  
In this paper we neglect its weak dependence on $N/Z_{QP}$ and treat 
it as a constant independent on $N/Z_{QP}$. 
Within our approximation, the resulting expression for 
the isobaric ratio Y($^{3}$H)/Y($^{3}$He) will be 

\begin{equation}
\ln(Y(^{3}\hbox{H})/Y(^{3}\hbox{He})) = \ln(K(T)) + (\mu _{n}-\mu _{p})/T 
\label{r3ln}
\end{equation}

\noindent 
where $ K(T) $ is a proportionality factor dependent on the temperature 
but independent of the N/Z ratio of the fragmenting system. According 
to the evaluation formulas for the free energy given in \cite{bondorf-smm}, 
one may expect the values of proportionality factor $ K(T) $ to be close 
to unity. 

\begin{figure}[!tbp]
\centering
\vspace{10mm}
\includegraphics[height=13cm]{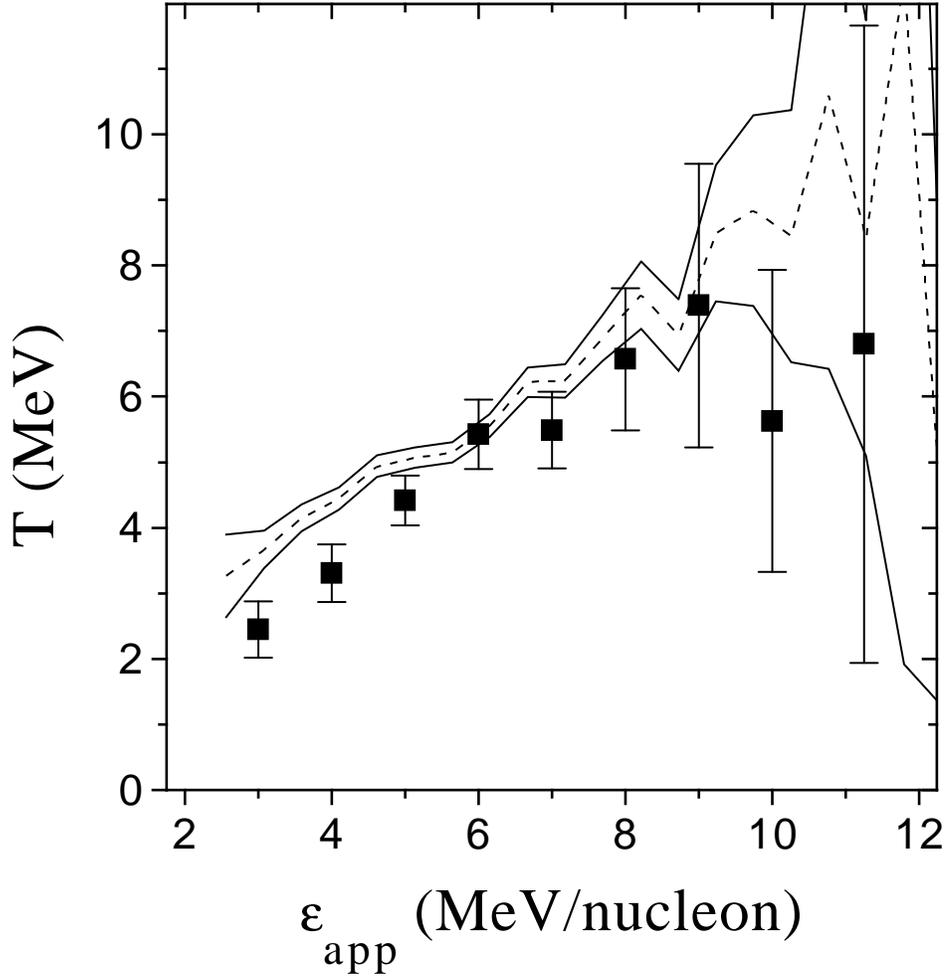}
\vspace{10mm}
\caption{ 
Dependence of the temperature $ T $, 
determined from the dependence of the isobaric ratio 
Y($^{3}$H)/Y($^{3}$He) on the N/Z ratio of 
the fragmenting quasiprojectile, on $\epsilon_{app}$ 
\hbox{( solid squares )}. Dashed and solid lines 
indicate the values and the statistical deviations of the temperature 
determined using the double isotope ratio method 
for the thermometer  \hbox{d,t/$^{3}$He,$^{4}$He}.  
}
\label{fig3}
\end{figure}

\begin{figure}[!tbp]
\centering
\vspace{10mm}
\includegraphics[height=13cm]{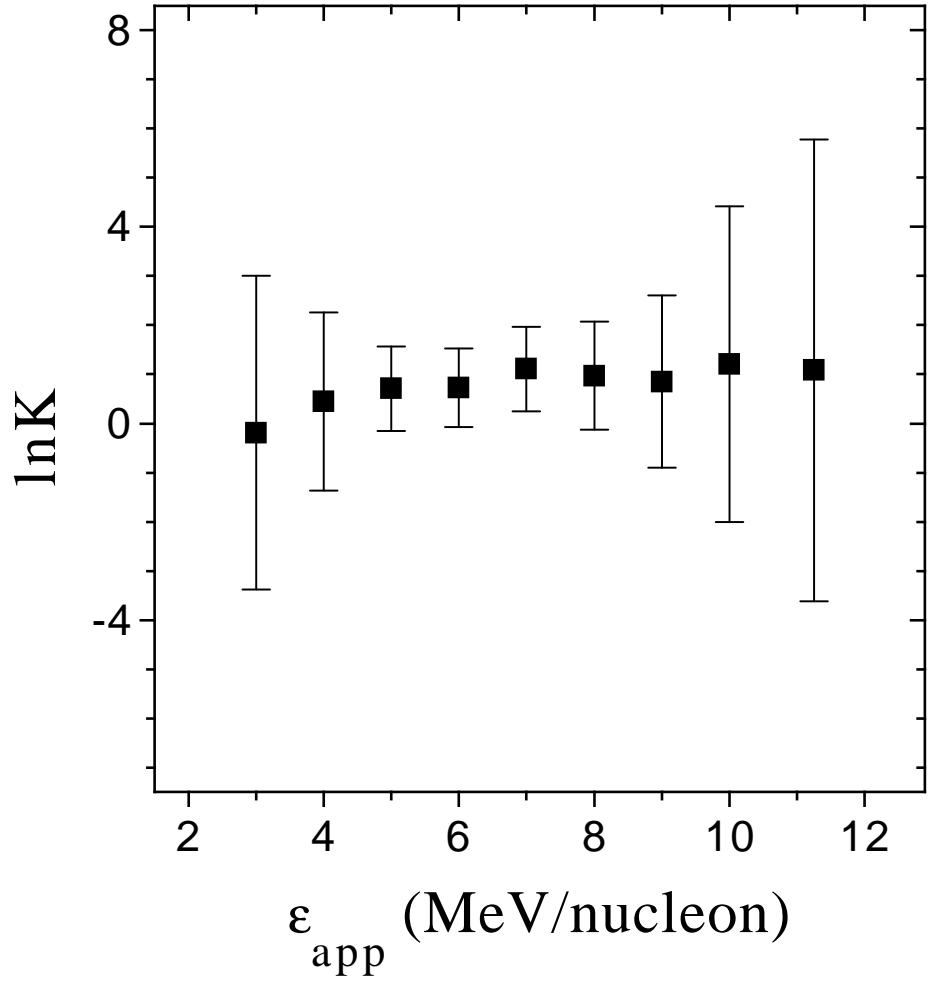}
\vspace{10mm}
\caption{ 
Dependence of $\ln(K(T))$, 
determined from the dependence of the isobaric ratio 
Y($^{3}$H)/Y($^{3}$He) on $N/Z_{QP}$, on $\epsilon_{app}$. 
}
\label{fig4}
\end{figure}

Since both ln(Y($^{3}$H)/Y($^{3}$He)) and 
$ \mu _{n}-\mu _{p} $ depend linearly on $ N/Z_{QP} $ ( as may be 
seen from Fig. \ref{fig2} and equation \ref{bnbp} ) 
and since we choose the value of $ \ln (K(T)) $ to be a constant, 
independent of $ N/Z_{QP} $, the expressions on both sides of 
equation \ref{r3ln} are in fact first order polynomials of $ N/Z_{QP} $. 
Then the solution is obvious, constants and coefficients of first order 
\hbox{( slopes )} should be equal on both sides of the equation. 
This allows a determination not only of the temperature ( as a ratio of 
the slopes ) but also of ln($K(T)$) from the comparison of the zero order 
coefficients \hbox{( constants )} using the extracted temperature. 
The resulting values of $ T $ and $ \ln (K(T)) $ are given 
in Fig. \ref{fig3} and Fig. \ref{fig4} for the different quasiprojectile 
excitation energy bins. 
The numerical values of $ T $ and $ \ln (K(T)) $ are given in the 
Table \ref{table1} along with the parameters of 
the linear fits of ln(Y($^{3}$H)/Y($^{3}$He)) given in Fig. \ref{fig2}. 
The dependence of the temperature on the excitation energy ( caloric 
curve ) given in Fig. \ref{fig3} ( solid squares ) 
is compared to the experimental caloric curve 
obtained for the same set of the data by the double isotope 
ratio method \cite{doublerat1,doublerat2} for the thermometer 
\hbox{d,t/$^{3}$He,$^{4}$He} \hbox{( the }dashed line indicates 
the double isotope ratio temperature and the solid lines indicate 
the statistical \hbox{errors ).} The formula 

\begin{equation}
T_{\rm d,t/^{3}He,^{4}He} = 
14.31 {\rm MeV}/\ln(1.59\frac{Y({^{2}{\rm H}})/Y({^{3}{\rm H})}}
{Y({^{3}{\rm He}})/Y({^{4}{\rm He}})})
\label{tdbliso}
\end{equation}

was used for the calculations. The agreement between the two plots is 
reasonable. Both methods give the value of the temperature between 5 and 7 MeV 
for the apparent excitation energies above 5 MeV/nucleon. 
Similar values of the temperature are reported by other experimental studies 
\cite{calor1,calor2,calor3,calor4}. The values of the temperature 
predicted by the SMM \cite{bondorf-smm} are also in the same range. 
The values of $ \ln (K(T)) $ are between 0 and 1 and suggest the values
of $ K(T) $ between 1 and 3, which are in good agreement with the values 
of the SMM. The dependence of $ \ln (K(T)) $ on the temperature 
is quite weak, which is also consistent with the expectations. 
Large relative errors of the $ \ln (K(T)) $ values are 
caused by the subtraction of numbers of similar magnitude. 
The agreement between temperatures determined by two 
different methods shows that the assumptions made for $ \mu _{n}-\mu _{p} $ 
reflect the physical trends which take place in the freeze-out configuration. 

Since the neutron emission from the quasiprojectile is not included 
in the data presented here, we tried to estimate its influence. 
Using a backtracing procedure described in 
\cite{veselsky}, we estimated the initial values of the excitation energy 
of a hot quasiprojectile at the stage where it also includes the neutrons 
that are later emitted. The mean value of the excitation energy taken 
away by a neutron emission ranges from 0.5 MeV/nucleon for the 
highest bin of $\epsilon_{app}$ to 1.5 MeV/nucleon for the lowest bin. 
This applies to both methods of the temperature determination presented here 
and implies that the events represent a still broader range of the 
initial excitation energies of the quasiprojectile. 
Since the multiplicity of the charged fragments decreases 
with decreasing $\epsilon_{app}$, the decrease of the determined 
temperatures in the lowest bins of $\epsilon_{app}$ may 
be a signature of the onset of a low energy deexcitation mode 
where several neutrons and light charged particles are emitted prior 
to the breakup of the partially cooled residue into two massive fragments 
( such a mass distribution is observed in the channels with 3 and 4 charged 
fragments ). The temperatures determined thus become mean values 
representing the range of temperatures at which hydrogen and helium 
isotopes are emitted during the deexcitation cascade. 
The use of $^{3}$He for the evaluation of the isobaric ratio 
Y($^{3}$H)/Y($^{3}$He) could be affected by the so-called 
$^{3}$He-puzzle ( see \cite{neubert} ), nevertheless it appears that the 
$^{3}$He-anomaly is not relevant for peripheral collisions induced 
by light projectile nuclei. 

%\section{Summary}

We presented the dependence of the isobaric ratio Y($^{3}$H)/Y($^{3}$He) on 
$N/Z_{QP}$ for the reaction of a $^{28}$Si beam on $^{112,124}$Sn targets 
at two different projectile energies 30 and 50 MeV/nucleon. 
We demonstrated a linear dependence of ln(Y($^{3}$H)/Y($^{3}$He)) 
on $N/Z_{QP}$ and the dependence of the slope on the 
excitation energy of the quasiprojectile. Using the model assumptions of the 
statistical multifragmentation model, we demonstrated the possibility 
of relating the slope of the dependence of ln(Y($^{3}$H)/Y($^{3}$He)) on 
$N/Z_{QP}$ to the temperature at a given excitation energy. 
We constructed a caloric curve and demonstrated its equivalence to the 
results of the double isotope ratio method. Our results show that 
the temperatures extracted from the isospin dependence 
of Y($^{3}$H)/Y($^{3}$He) are consistent with the temperature of 
the freeze-out configuration where thermal equilibration occurs. 
The approximation used for the evaluation of the difference of neutron 
and proton chemical potentials is consistent with significant changes 
of densities of neutrons and protons during multifragmentation.    

%\begin{acknowledgments}

The authors wish to thank the Cyclotron Institute staff for the excellent
beam quality. This work was supported 
in part by the NSF through grant No. PHY-9457376, 
the Robert A. Welch Foundation through grant No. A-1266, and 
the Department of Energy through grant No. DE-FG03-93ER40773. 
M. V. was partially supported through grant VEGA-2/5121/98.

%\end{acknowledgments}

%Since the method seems to
%be promising we expect to carry on the studies with heavier projectiles 
%at the NIMROD \cite{NIMROD} facility of the Cyclotron Institute 
%at Texas A\&M University. 

\newpage

\newpage

\newpage

%\begin{table}[tbp]
\begin{table}[!tbp]
\caption{ Parameters of the exponential fit 
$\frac{Y(^{3}\hbox{H})}{Y(^{3}\hbox{He})}= \exp(a_{1}+a_{2}(\frac{N}{Z})_{QP})$ 
for the different bins of the apparent excitation energy $\epsilon_{app}$ 
and the determined values of the temperature $ T $ and $ \ln (K(T)) $ for 
the same bins.  }
\label{table1}

\vspace{0.9cm}
{\centering \begin{tabular}{ccccc}
\hline 
\hline 
$ \epsilon_{app} $, MeV/nucleon&
$ a_{1} $&
$ a_{2} $&
$ T $, MeV&
$ \ln (K(T)) $\\
\hline 
2.5-3.5&
-17.895$ \pm  $2.170&
15.719$ \pm  $2.007&
2.45$ \pm  $0.43&
-0.19$ \pm  $3.19\\
3.5-4.5&
-12.665$ \pm  $1.175&
11.642$ \pm  $1.160&
3.31$ \pm  $0.44&
0.45$ \pm  $1.81\\
4.5-5.5&
-9.124$ \pm  $0.517&
8.730$ \pm  $0.538&
4.42$ \pm  $0.38&
0.71$ \pm  $0.86\\
5.5-6.5&
-7.267$ \pm  $0.483&
7.099$ \pm  $0.515&
5.43$ \pm  $0.53&
0.73$ \pm  $0.80\\
6.5-7.5&
-6.808$ \pm  $0.518&
7.031$ \pm  $0.564&
5.49$ \pm  $0.58&
1.11$ \pm  $0.86\\
7.5-8.5&
-5.634$ \pm  $0.682&
5.868$ \pm  $0.744&
6.57$ \pm  $1.08&
0.97$ \pm  $1.10\\
8.5-9.5&
-5.023$ \pm  $1.084&
5.218$ \pm  $1.210&
7.39$ \pm  $2.16&
0.85$ \pm  $1.75\\
9.5-10.5&
-6.499$ \pm  $2.020&
6.846$ \pm  $2.202&
5.63$ \pm  $2.30&
1.21$ \pm  $3.21\\
10.5-12.5&
-5.323$ \pm  $2.931&
5.689$ \pm  $3.241&
6.80$ \pm  $4.86&
1.08$ \pm  $4.69\\
\hline 
\hline 
\end{tabular}\par}
\end{table}

\end{document}